\documentclass[10pt,letterpaper]{article}
\usepackage{ol2}

\usepackage{graphicx}
\DeclareGraphicsExtensions{.pdf}

\begin{document}

\title{Observation of electro-activated localized structures in broad area VCSELs}%


\author{J. Parravicini$^{1}$, M. Brambilla$^{2,3}$, L. Columbo$^{3,4}$, F. Prati$^{4,5}$, C. Rizza$^{4,6}$, G. Tissoni$^{7}$,  A. J. Agranat$^{8}$, and E. DelRe$^{1 *}$}%

\address{$^1$Dipartimento di Fisica and IPCF-CNR, ``Sapienza'' Universit\`{a} di Roma,
I-00185 Roma, Italy}%

\address{$^2$Dipartimento Interateneo di Fisica, Universit\`{a} e Politecnico di Bari, I-70126 Bari, Italy}

\address{$^3$CNR-IFN, Bari I-70126, Italy}

\address{$^4$Dipartimento di Scienza \& Alta Tecnologia, Universit\`{a} dell'Insubria, Como I-22100, Italy}

\address{$^5$CNISM, Research Unit of Como, Como I-22100, Italy}

\address{$^6$CNR-SPIN, Coppito L'Aquila I-67100, Italy}

\address{$^7$INLN, CNRS, Universit\'{e} de Nice Sophia Antipolis, Valbonne F-06560, France}

\address{$^8$Applied Physics Department, Hebrew University of Jerusalem, IL-91904 Israel}%

\email{*eugenio.delre@uniroma1.it}

%
%
%
%
%

%
%

\begin{abstract}
\noindent
We demonstrate experimentally the electro-activation of a localized optical structure in a coherently driven broad-area vertical-cavity surface-emitting laser (VCSEL) operated below threshold. Control is achieved by electro-optically steering a writing beam through a pre-programmable switch based on a photorefractive funnel waveguide.
\end{abstract}


\section{Introduction and motivation}
One of the main hurdles of photonic technology is the ability to store temporarily information directly from an optical stream \cite{Heinze2013}. An ideal platform is that of VCSELs that, when externally driven by a coherent laser field and operated below threshold, can support stable localized light structures in response to an appropriate writing optical signal \cite{Akhmediev2008,Barbay2011}. Not only are VCSELs technologically integrated with standard electronics, so that they can form a valid bridge between an electronic motherboard and a guided or free-air optical network, but they can be built in a so-called ``broad area'' format, so that the encoded spatial structure only pervades a small micron-sized transverse portion of the 100+ $\mu$m cavity \cite{Barbay2011,Spinelli1998}. This allows, in specific conditions, a dynamic formation of localized structures known as cavity solitons \cite{Lugiato2008,Barland2002,Genevet2008} that can move and interact \cite{Firth1996,Maggipinto2000,Averlant2014}. Since the structures are optically generated, an important issue is how to achieve their rapid control through an electrical signal. Fundamentally, this is an issue of spatial light modulation, so that standard techniques based on micro-arrays of liquid-crystals or mirrors are applicable \cite{Pedaci2006}. In turn, these techniques are burdened by intrinsic limitations associated to a millisecond response time. An alternative scheme is  to make use of acousto-optic modulation  \cite{Hachair2005} and, ideally,  electro-optics, where response time down to and below the nanosecond scale is readily achieved.

Recently, Columbo et al. \cite{Columbo2012,Columbo2014}  theoretically investigated the idea of steering localized structures in a VCSEL through soliton electro-activation \cite{DelRe2000,DelRe2002}.  In fact, spatial solitons in photorefractive crystals \cite{DelRe2009} can embed inside the electro-optic sample optical waveguides \cite{Shih1996} that can be arranged into arrays \cite{Petter2003}.  When this is achieved inside a paraelectric sample, the waveguides can be switched on and off with nanosecond response times \cite{Sapiens2009}, forming a miniaturized array of electro-optic switches \cite{Dercole2004,Asaro2005}.  An even more versatile electro-activated waveguide is achieved using so-called electro-activated funnel waveguides \cite{DelRe2007,Pierangelo2009,DelRe2012}, where the  limits to photorefractive spatial soliton waveguides associated to bending \cite{Carvalho1995,Krolikowski1996,DelRe2005}, circular symmetry \cite{DelRe2004}, and time stability \cite{Fressengeas1998,DelRe2006}, are absent.

In this paper we demonstrate the switching on of an 11 $\mu$m (Full-Width-at-Half-Maximum, FWHM) localized optical structure inside a broad area (200 $\mu$m-diameter) VCSEL cavity through an electro-activated funnel waveguide structure in an electrically biased sample of potassium-lithium-tantalate-niobate (KLTN).  We are also able to show that the localized structure is bistable, that is, it remains on even if the writing signal is eventually switched off, and that a single localized structure can be addressed independently, that is, without exciting other similar localized structures in the cavity.


\section{Experimental setup}

\begin{figure}
\centering
\includegraphics[width=0.99\columnwidth]{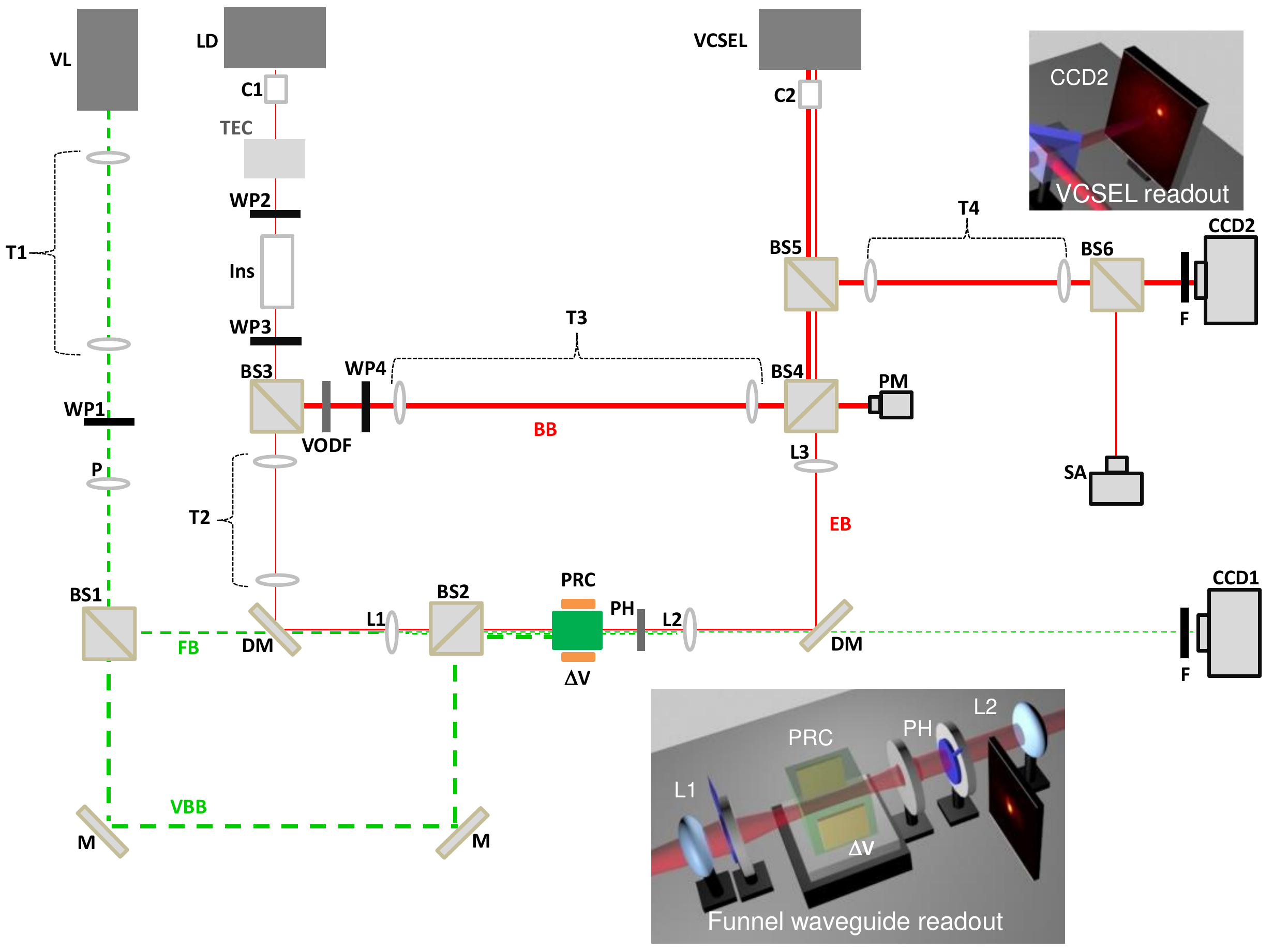}
\caption{Detailed scheme of experimental setup. VL: visible-wavelength ($\lambda=$532 nm) cw laser. T1, T2, T3, T4: adjustable optical beam-expanders. WP1, WP2, WP3, WP4: $\lambda/2$ waveplates. P: polarizer. BS1, BS2, BS3, BS4, BS5, BS6: beam-splitters. LD: infrared laser diode. C1, C2: collimators. TEC: Tunable external cavity in Littman-Metcalf configuration. Ins: optical insulator. L1, L2, L3: lenses. VODF: variable optical density filter. M: mirrors. DM: dichroic mirrors. PH: iris. PRC: electrically-driven photorefractive crystal. VCSEL: broad-area VCSEL cavity. F: neutral density filters. CCD1, CCD2: CCD cameras. PM: power meter. SA: spectrum analyzer. FB: focused beam ($\lambda=$532 nm). VBB: longitudinally-broad visible background beam ($\lambda=$532 nm). BB: longitudinally broad infrared ($\lambda=$977 nm) background beam. EB: focused near-infrared ($\lambda=$977 nm) exciting beam.}%
\label{FigureSetup}
\end{figure}

The detailed  setup is reported in the scheme of Fig. \ref{FigureSetup}. The apparatus is the merging of two stages: a standard setup for photorefractive soliton generation (as reported e.g. in \cite{DelRe2009}) and a setup for semiconductor-cavity injection and inspection (see e.g. \cite{Lugiato2008,Barland2002}).

In the first stage, a cw visible laser beam ($\lambda= 532$ nm, 200-400 mW) is first expanded (T1), then, after polarization and power tuning (WP1 and P), split in two branches (BS1). The first one (FB) is focused (FWHM of $\simeq$ 12 $\mu$m, with a power of 1-3 $\mu$W), by means of the adjustable L1 lens, on the input facet of a photorefractive crystal (PRC) to produce either a photorefractive soliton or a waveguide, while the other (VBB) propagates as a plane-wave until it is recombined with FB by the beam-splitter BS2 and illuminates the whole input facet of PRC as a background. The intensity profile of the light at the ouput of the PRC is recorded on a CCD camera (CCD1) through an adjustable imaging lens (L2).

In the second stage, an infrared laser beam is generated by a semiconductor laser diode (LD), collimated (C1) and filtered by a tunable Littman-Metcalf cavity (TEC) to  $\lambda\simeq$977 nm. It then passes through an optical insulator (Ins) and it is split in two branches by a polarizing beam-splitter (BS3) (the power in the two ouput channels can be balanced through WP3). One beam (the exciting beam EB) becomes collinear with FB through a dichroic mirror (DM) and, using T2 and L1, it is focused onto the input facet of PRC. The second beam (the background beam BB) is expanded by T3 and arrives onto BS4, where it recombines with EB. The collimator C2 injects both EB and BB (with the same polarization) in a broad-area (diameter of 200 $\mu$m) VCSEL to excite the formation of localized structures.

\begin{figure}
\centering
\includegraphics[width=0.6\columnwidth]{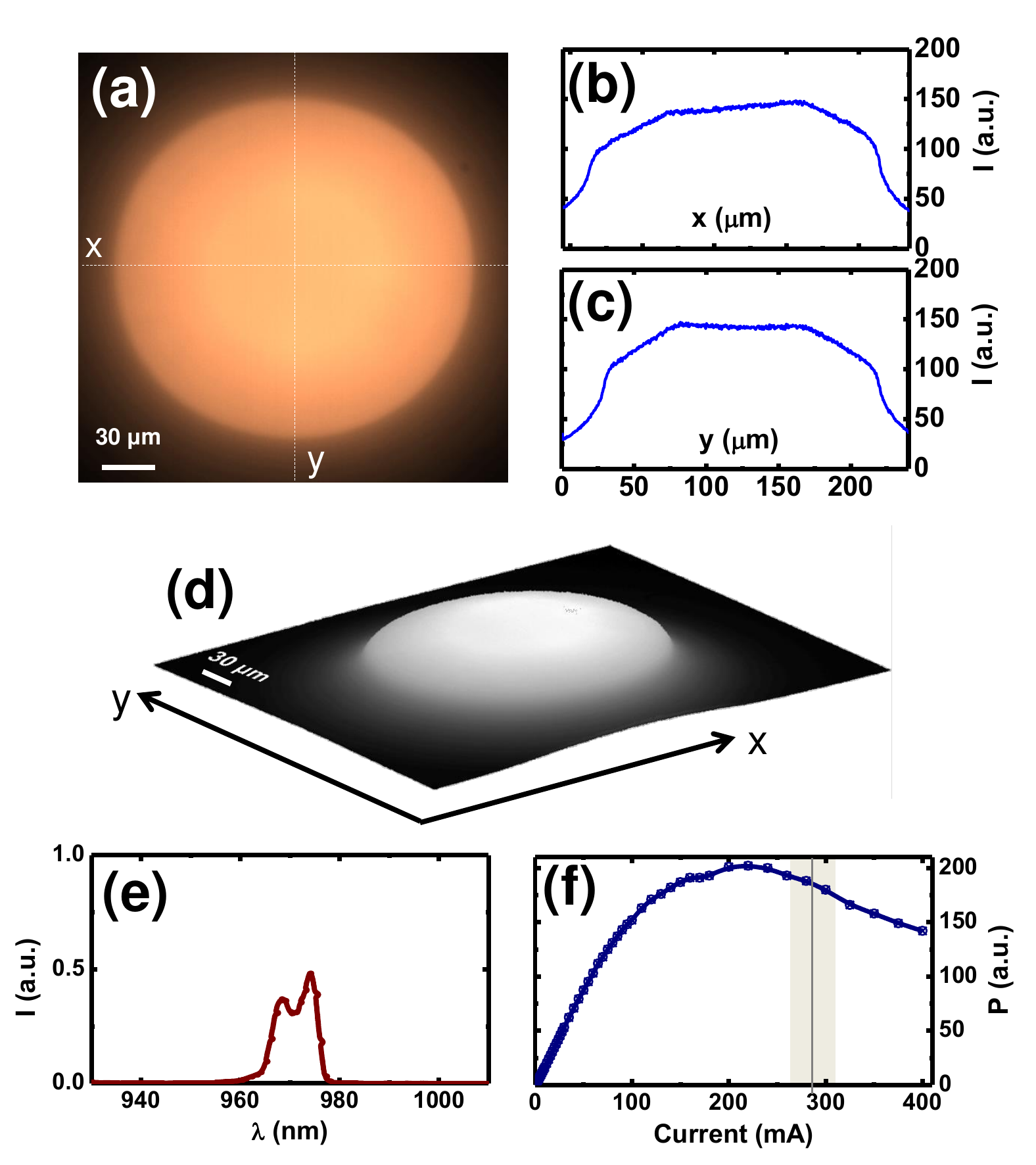}
\caption{(a) Intensity distribution of the VCSEL emission (kept below threshold) with intensity profiles in $x$ (b) and $y$ (c) directions. (d) 3D emission profile of the VCSEL. (e) typical spectrum of the VCSEL below-threshold (FWHM$\simeq$18.3 nm) and (f) current-power characteristic curve: saturation effects below-threshold occur due to thermal rollover, as is typical in continuous-current-supplied broad-area VCSELs \cite{Michalzik2003}. Shaded region in (f) indicates the range where localized structures are observed.}%
\label{FigureVCSEL}
\end{figure}

The VCSEL is maintained below threshold as in Ref.\cite{Columbo2012} and exhibits a broad emission spectrum, from 965-978 nm (see Fig. \ref{FigureVCSEL}). EB, which exits from from PRC having FWHM $\simeq$ 25 $\mu$m, passes through the iris PH and is imaged onto the plane of the VCSEL, where it has a FWHM of $\simeq$ 10 $\mu$m (with a power of 20-40 $\mu$W). The BB, instead, enters the VCSEL with a FWHM of 280-300 $\mu$m and a power of 10-50 mW. The light emitted by the VCSEL is then steered to the detection branch trough a 90:10 beam-splitter (BS5): finally it is focused onto a camera (CCD2) by T4 and a part of the power is sent to a spectrum analyzer SA (with a resolution of $\Delta\lambda\simeq$ 0.05 nm). The relative phase difference between BB (near plane-wave) and EB (focused Gaussian beam) is inspected through the left branch of BS5 and, in our specific conditions, is maintained constant.

\begin{figure}
\centering
\includegraphics[width=0.7\columnwidth]{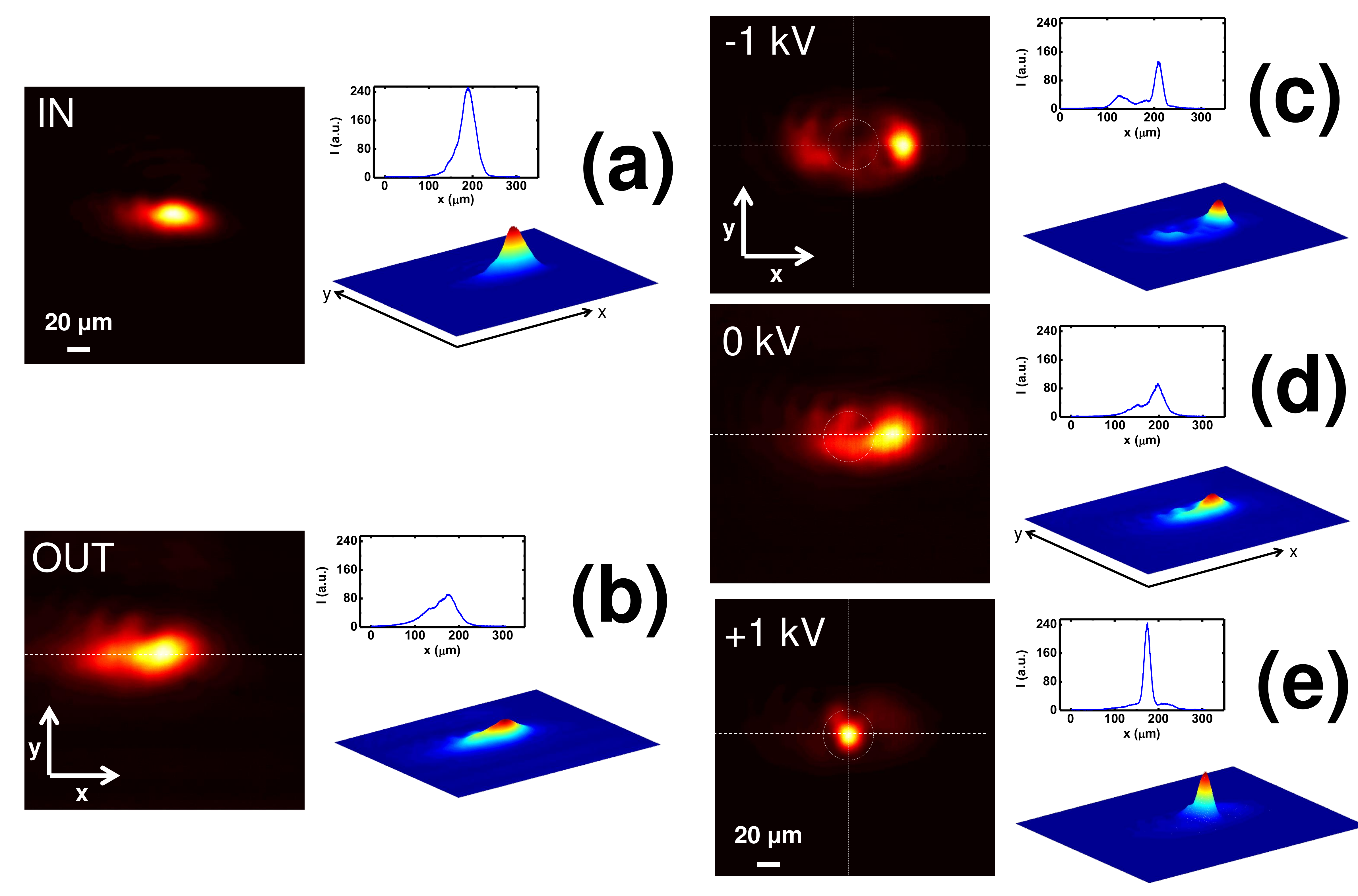}
\caption{IR ($\lambda=$ 977 nm) intensity distribution of the beam at the input facet (a) and output facet (b) (after 2.7 mm of standard linear propagation) of the crystal. Electro-driven guiding of the IR beam in a double-funnel waveguide previously written through a visible-light laser beam ($\lambda=$ 532 nm): intensity distribution at the output facet of the crystal for an applied voltage of -1 kV (c) (note the antiguiding central region and the guiding lateral lobes), 0 V (d) and +1 kV (e) (with the central guiding region fully activated). The central circle in (c), (d), (e) indicates the region, selected by the iris, which is imaged in the plane of the VCSEL.}%
\label{FigureGuiding}
\end{figure}

The electro-optic switch is achieved using a double-funnel photorefractive waveguide \cite{Pierangelo2009,DelRe2012} in a zero-cut 2.7$^{(c)}$ $\times$ 9.9$^{(b)}$ $\times$ 2.4$^{(a)}$ mm Cu-doped sample of KLTN with a composition   K$_{1-y}$ Li$_{y}$Ta$_{1-x}$Nb$_{x}$O$_{3}$, where $x=0.003$, $y=0.35$  \cite{Agranat1992}. The waveguide is previously written at 532 nm and is switched on or off during readout at infrared wavelengths by means of a power-supply delivering a time-constant voltage in the range from $\pm$1 kV, which corresponds to a field in the range of $\pm$3.7 kV/cm (see Fig. \ref{FigureGuiding} below).

\begin{figure}
\centering
\includegraphics[width=1.1\columnwidth]{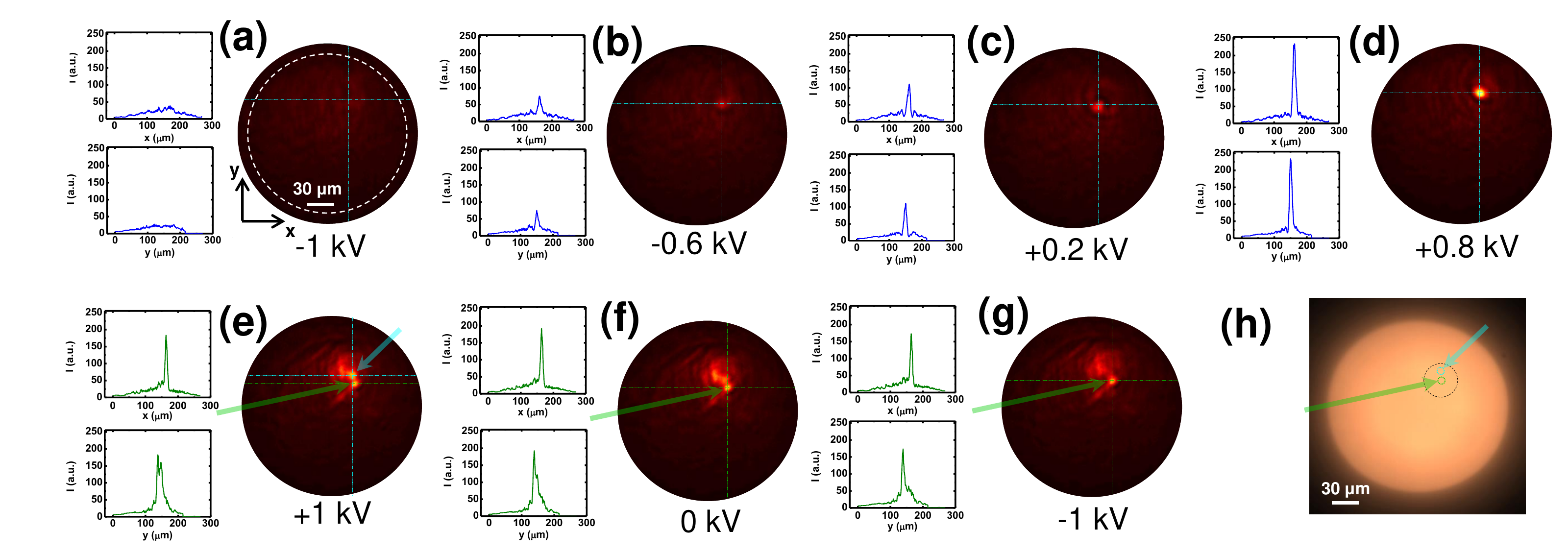}
\caption{(a), (b), (c), (d), (e), (f), (g): VCSEL emission intensity pattern in $x-y$ plane (right) and relative intensity profiles centered in the dashed cross (left); blue crosses are centered in EB spot, while green crosses are centered in excited localized spot. (a), (b), (c), (d): emission at exciting spot EB intensities for respectively $\Delta V=$ -1, -0.6, +0.2, +0.8 kV applied voltage on the crystal (intensity profiles are centered at EB spot); (e), (f), (g): VCSEL emission after the structure switches on (intensity profiles are centered in excited localized structure) for $\Delta V=$ +1, 0, -1 kV. For comparison, we indicate in (h) the position of the EB spot (green circle) and that of the localized structure (blue circle), along with the region in which activation occurs (dashed circle) relative to the transverse intensity profile of the VCSEL (without no injected beams, as in Fig. \ref{FigureVCSEL}a).}%
\label{Modulation}
\end{figure}


\section{Results}
In Fig. \ref{FigureGuiding} we demonstrate the electro-modulation of the 977 nm signal from the diode laser through the electro-activated funnel waveguide. In a first step, the double-funnel guide is written with a 532 nm laser (writing phase). Then, in the read-out phase, a focused infrared beam (EB) is launched into the waveguide and its propagation is electro-optically driven by means of the electroactivation, that is, by the nonlinear combination of the applied static field and the previously imprinted space-charge field caused by the quadratic electro-optic response \cite{Pierangelo2009}.

\begin{figure}
\centering
\includegraphics[width=0.45\columnwidth]{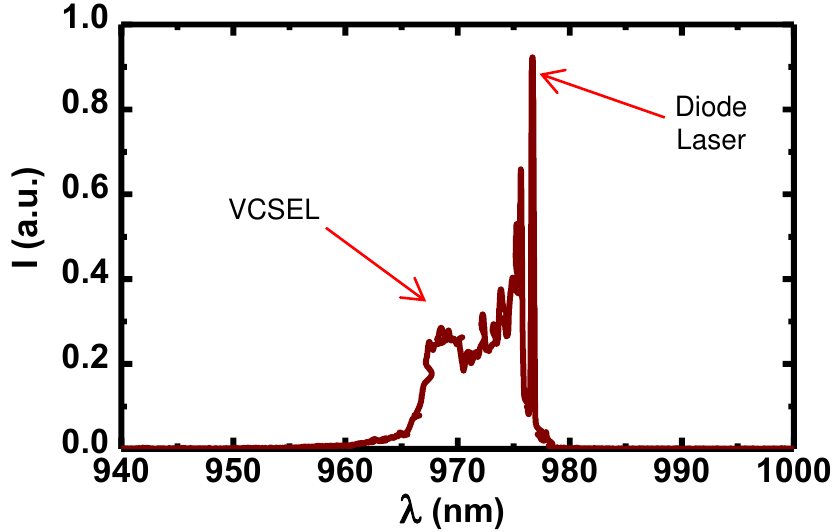}
\caption{Spectrum of VCSEL emission (below threshold, Fig. \ref{FigureVCSEL}e) with superimposed spectrum of injected beams (FWHM$\simeq$ 0.43 nm) from the diode laser (LD) in experimental conditions.}%
\label{FigurePath}
\end{figure}

\noindent In Fig. \ref{Modulation} we demonstrate the electro-activation of a localized structure in the VCSEL and its hysteresis.  As reported in Figs. \ref{Modulation}a-d, when the field delivered to the PRC is increased from -1 kV to +0.8 kV, the VCSEL begins emitting where the EB impinges (crossing of the blue-dashed lines).  The emitted peak intensity  increases by 20 times.  When +1 kV is applied, something qualitatively different occurs: a localized structure abruptly forms (the temporal resolution is limited by the single figure frame capture of the CCD, around 100 ms), as shown in Fig. \ref{Modulation}e.  The localized structure forms pinned to a fixed position (green dashed line crossing) in the VCSEL that does not need to coincide with the position of the EB, as illustrated in Fig. \ref{Modulation}h.  Once this transition has occurred, the pinned localized structure persists unchanged even if the voltage delivered to the PRC is decreased to -1 kV, in which case the VCSEL emission at the original position of the EB is strongly attenuated (Figs. \ref{Modulation}f and \ref{Modulation}g).  The switch-on phenomenon is observed repositioning the EB throughout the region surrounding the position of the localized structure, illustrated with the dashed circle in Fig.\ref{Modulation}h, a region of approximately 30 $\mu$m diameter.  The switching is also dependent on the relative phase of EB and BB, which in turn depends on the actual applied voltage $\Delta V$.  In our reported instance, the relative phase is optimized during alignment so that switching occurs only when the voltage +1 kV is delivered.  Finally, the localized structure is switched off, erasing the information stored in the VCSEL, either by lowering the pump current or changing the mismatch of the driving coherent field.

The superimposed spectra of both VCSEL and diode beams (BB and EB) are shown in Fig. \ref{FigurePath}: the mismatch between the highest peak of VCSEL spectrum (Fig. \ref{FigureVCSEL}e) and the wavelength of the injected beam is $\Delta\lambda\simeq$ 1 nm. Note that, in agreement with theory, structures are observed when the injected laser wavelength is longer than the highest components of the VCSEL spectrum \cite{Spinelli1998}.


\section{Conclusion}
We have demonstrated the activation of a localized $\simeq$11 $\mu$m structure at 977 nm inside a 200 $\mu$m VCSEL through an electro-optically controlled writing beam. The activating beam is controlled using an electro-optic funnel waveguide embedded inside the volume of a paraelectric near-transition sample of KLTN. This system realizes the first, to our knowledge, hybrid all-optical device integrating propagative and dissipative soliton-like structures. Further developments include the demonstration of electro-deactivation of the localized structure and the encoding of a spatially complex array of electro-optically-driven independent spots. From an applicative perspective, our present result allows the electro-activation of localized structures through an intrinsically rapid response, down to the nanosecond scale.


\section*{Acknowledgements}
We thank Stephane Barland for helpful discussions and suggestions. This work was funded by FIRB grant PHOCOS-RBFR08E7VA.


\end{document}